%% file: root.tex
\documentclass[10pt]{IEEEtran}
\usepackage{mathtools}
\usepackage{lipsum}
\usepackage{graphicx}

\input{preambleT.tex}

\usepackage{blindtext, graphicx}   
\usepackage{amsmath}   
\usepackage[tc]{titlepic}   
\usepackage{amssymb}   
\usepackage{amsfonts}   
\usepackage{lipsum}   
\usepackage{float}   
\usepackage{cuted}     
\usepackage{amsthm}   
\newtheoremstyle{case}{}{}{}{}{}{:}{ }{}
\usepackage{enumitem}
\setlist[itemize]{leftmargin=*}
\setlist[description]{leftmargin=*}
\usepackage{subfigure}
\usepackage[margin=0.75in]{geometry}

\title{\vspace{0.7cm}\LARGE  \textbf{Dynamic Actuator Selection and Robust State-Feedback Control of Networked Soft Actuators}}
\author{Nafiseh Ebrahim$\text{i}^{*,**}$, Sebastian Nugroh$\text{o}^{\dagger}$, Ahmad F. Tah$\text{a}^{\dagger,\ddagger}$, Nikolaos Gatsi$\text{s}^{\dagger}$, Wei Ga$\text{o}^{*}$, Amir Jafar$\text{i}^{*,**}$ \vspace{-0.8cm}
	\thanks{$^\ddagger$Corresponding author. $\text{}^{*}$Department of Mechanical Engineering. $\text{}^\dagger$Department of Electrical and Computer Engineering. $\text{}^{**}$Advanced Robotic Manipulators (ARM) Lab.
	All authors are with the University of Texas at San Antonio, 1 UTSA Circle, San Antonio, TX 78249. Emails:
	\{nafiseh.ebrahimi, ahmad.taha, amir.jafari, nikolaos.gatsis, wei.gao\}@utsa.edu, sebastian.nugroho@my.utsa.edu.This material is based upon work supported by the National Science Foundation under Grant CMMI-DCSD-1728629.}}

\begin{document}

\maketitle
\thispagestyle{empty}
\pagestyle{empty}

\begin{abstract}
The design of robots that are light, soft, powerful is a grand challenge. Since they can easily adapt to dynamic environments, soft robotic systems have the potential of changing the status-quo of bulky robotics. A crucial component of soft robotics is a soft actuator that is activated by external stimuli to generate desired motions. Unfortunately, there is a lack of powerful soft actuators that operate through lightweight power sources. To that end, we recently designed a highly scalable, flexible, biocompatible Electromagnetic Soft Actuator (ESA). With ESAs, artificial muscles can be designed by integrating a network of ESAs. The main research gap addressed in this work is in the absence of system-theoretic understanding of the impact of the realtime control and actuator selection algorithms on the performance of networked soft-body actuators and ESAs. The objective of this paper is to establish a framework that guides the analysis and robust control of networked ESAs. A novel ESA is described, and a configuration of soft actuator matrix to resemble artificial muscle fiber is presented. A mathematical model which depicts the physical network is derived, considering the disturbances due to external forces and linearization errors as an integral part of this model. Then, a robust control and minimal actuator selection problem with logistic constraints and control input bounds is formulated, and tractable computational routines are proposed with numerical case studies.
\end{abstract}
\vspace{-0.25cm}
\section{Introduction, Brief Literature Review, Paper Contributions and Organization}

The emerging field of soft robotics represents the foundation of future robotic systems with plethora of applications in human-robot interaction, locomotion, and rehabilitation technologies~\cite{RN1}. A crucial component of soft robotics is a \textit{soft actuator} that is activated to generate desired motions. Unfortunately, there is a lack of actuators in rehabilitation applications that are portable, adoptable to different joint sizes, while still matching the performance of the mammalian muscles in terms of response time and output power-to-size ratio. Consequently, the next generation of soft actuators that achieve assistive functions raises challenges to actuation design and analysis, materials engineering, modeling, and realtime robust control and optimization---the main theme of the proposed research. Motivated by these challenges we have recently designed a bio-inspired highly scalable, flexible, biocompatible Electromagnetic Soft Actuator (ESA)~\cite{SoRo} which we explain next. 
\begin{figure}[t]
	\vspace{-0.14cm}
	\centering
	\includegraphics[scale=0.25]{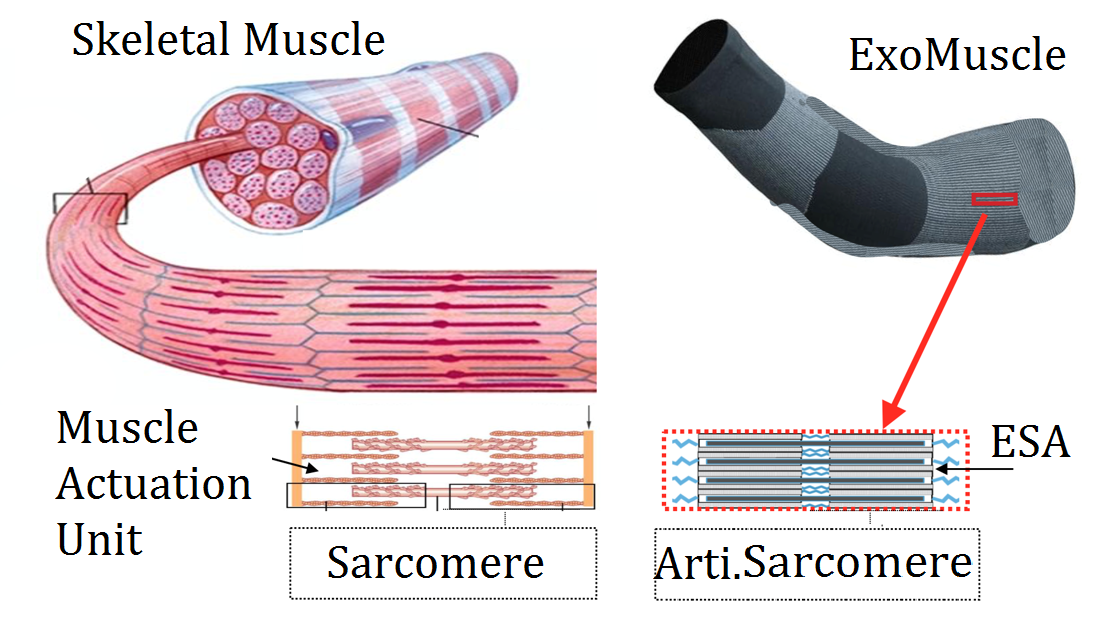}	
	\caption{ExoMuscles: the future of artificial muscles.}
	\label{fig:musc}
		\vspace{-0.33cm}	
\end{figure} 

As shown in Fig.~\ref{fig:musc}, a human skeletal muscle with bundles of sarcomeres (composed by actin and myosin) behaves like a network of soft actuators. A network of ESA can be integrated inside an artificial muscle (coined as {\textit{ExoMuscle}}) warped around the joints as active braces, mimicking a human muscle. Notably, it is analytically and experimentally confirmed that by scaling down the size of ESA, the ratio between resulting force to cross-section area  (F/CSA) of ESA increases~\cite{SoRo}. It is essential in obtaining large forces over a relatively small space using ESAs.

Latest relevant studies focus either on the design of soft material or the construction of soft actuators into artificial muscles. The main research gap lies in the absence of system-theoretic understanding of the impact of  realtime control algorithms on the performance of networked soft-body actuators in general, and ESA networks in specific. {The objective of this paper is to establish a framework that guide the design, analysis, and robust control of ESA networks forming artificial muscles under uncertainty. }


{Series elastic} and {variable stiffness actuators} (SEA/VSAs) are implemented and designed because of their ability to minimize large forces due to shocks, to safely interact with the user, and their ability to store and release energy in passive elastic elements \cite{RN26,RN22,RN32}. However, they are made of intrinsically rigid components. As a result, they are bulky. 
Actuators based on {shape memory alloys} have highly nonlinear behavior, low energy efficiency and low response speed \cite{RN34}, despite their advantages that include high power to weight ratio, mechanism simplicity, silent actuation, and low driving voltage. 
{Pneumatic artificial muscles} \cite{RN40,RN30} have been vastly used in robotics. However, they require stationary power sources and accessories such as air pump and valves.
{Dielectric elastomer actuators} \cite{RN46,RN48} are popularly referred to as artificial muscles because of their actuation speed, low density, and silent operation.  Unfortunately, they demand high operating voltages---preventing their operation with on-board batteries. 


Another dimension for successful operation of soft robotics in general, and artificial muscles in specific, is the realtime transient control problem which refers to the optimal current injections for each actuator in an networked system---as well as determining the subset of actuators to be activated given a desired motion and maximum input current. To that end, the problem of (a) simultaneously obtaining a minimal actuator subset while (b) designing localized, robust control laws for activated actuators is needed. Unfortunately, this routine of actuator selection and robust control is known to be a combinatorial, NP-hard problem. To address this challenge, various quantitative notions of network controllability and observability have been used~\cite{ruths2014control,pequito2016,pequito2017robust} and different techniques that are based on heuristics or simple linear time-invariant dynamics are presented in~\cite{summers2014submodularity,Dhingra2014,Chanekar2017}. 

Little is known about the interplay between actuator selection and the robust control problems with disturbances when maximum input bounds are imposed---for generic dynamic systems as well as for networked soft-body actuators. In this paper, we focus on addressing this gap which transcends ExoMuscles, and can be applied to various soft robotics. The paper contributions and organization are as follows.
\begin{itemize}
	\item The novel ESA and its properties are described, and a configuration of soft actuator matrix to resemble artificial muscle fiber is presented  (Section~\ref{sec:ESAs}). This is followed by the derivation of a physical network model of soft actuators. Disturbances due to external forces and linearization errors are considered as an essential part of the model (Section~\ref{sec:ModelingSS}).
	\item A robust control and minimal actuator selection problem formulation with actuator constraints and maximum input voltage bounds is investigated in Section~\ref{sec:RobustControl}. This formulation can be written as a nonconvex optimization problem with mixed-integer nonlinear matrix inequalities. To address the computational difficulties in solving this routine, tractable routines are explored in Section~\ref{sec:EfficientMethods} and~\ref{sec:BM}.
	\item Numerical tests are presented in Section~\ref{sec:numtests} showing the performance of the relaxations. The results suggest that network of ESAs can be robustly controlled with a subset of available actuators, while still yielding reasonable energy and stability performance. 
\end{itemize}
\section{Soft Actuator and System Configuration}~\label{sec:ESAs}
 \vspace{-1cm}
 
\subsection{Electromagnetic Soft Actuator (ESA) Components}
\begin{figure} [h]
\begin{center}
\includegraphics[scale=0.45]{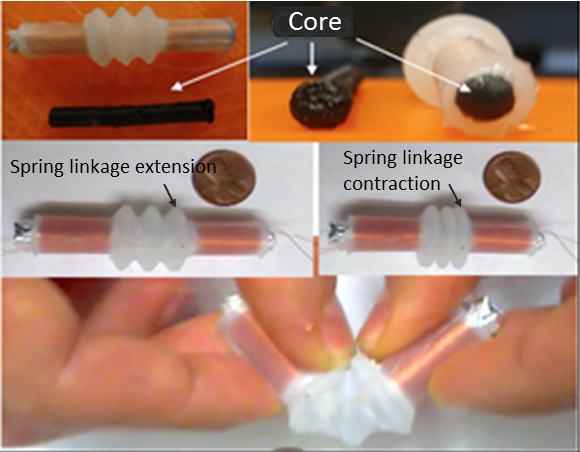}
\caption{Intrinsically soft electromagnetic actuator composed of two wire coils located on both sides of silicone spring linkage and a soft silicone-ferromagnetic core inside the coils having a clearance fit respect to them. }
\label{fig:esa}
\end{center}
\vspace{-0.5cm}
\end{figure}

Fig.~\ref{fig:esa} shows the electromagnetic soft actuator we recently designed~\cite{SoRo} and its components. It is made of biocompatible and magnetically permeable soft silicone that surrounds the copper wire coils and a soft silicone-ferromagnetic core.

This actuator is made mostly of silicone rubber so that it can have low stiffness. The major components of the soft actuator include helical coil, soft silicone ferromagnetic core, inner layer, spring linkage and outer layer. The helical coils are made of 100 turn of 34 American wire gauge copper. The outer layer of ESA which is included to shield and boost the resultant magnetic field consists of a mixture of $40\%$ iron oxide and $60\%$ silicone rubber. This part is included to make a layer of electromagnetic suspension to strengthen the generated field and increase the force.

\subsection{Electromagnetic Soft Actuator Test Setup}
\begin{figure} [t]
\centering \includegraphics[scale=0.536]{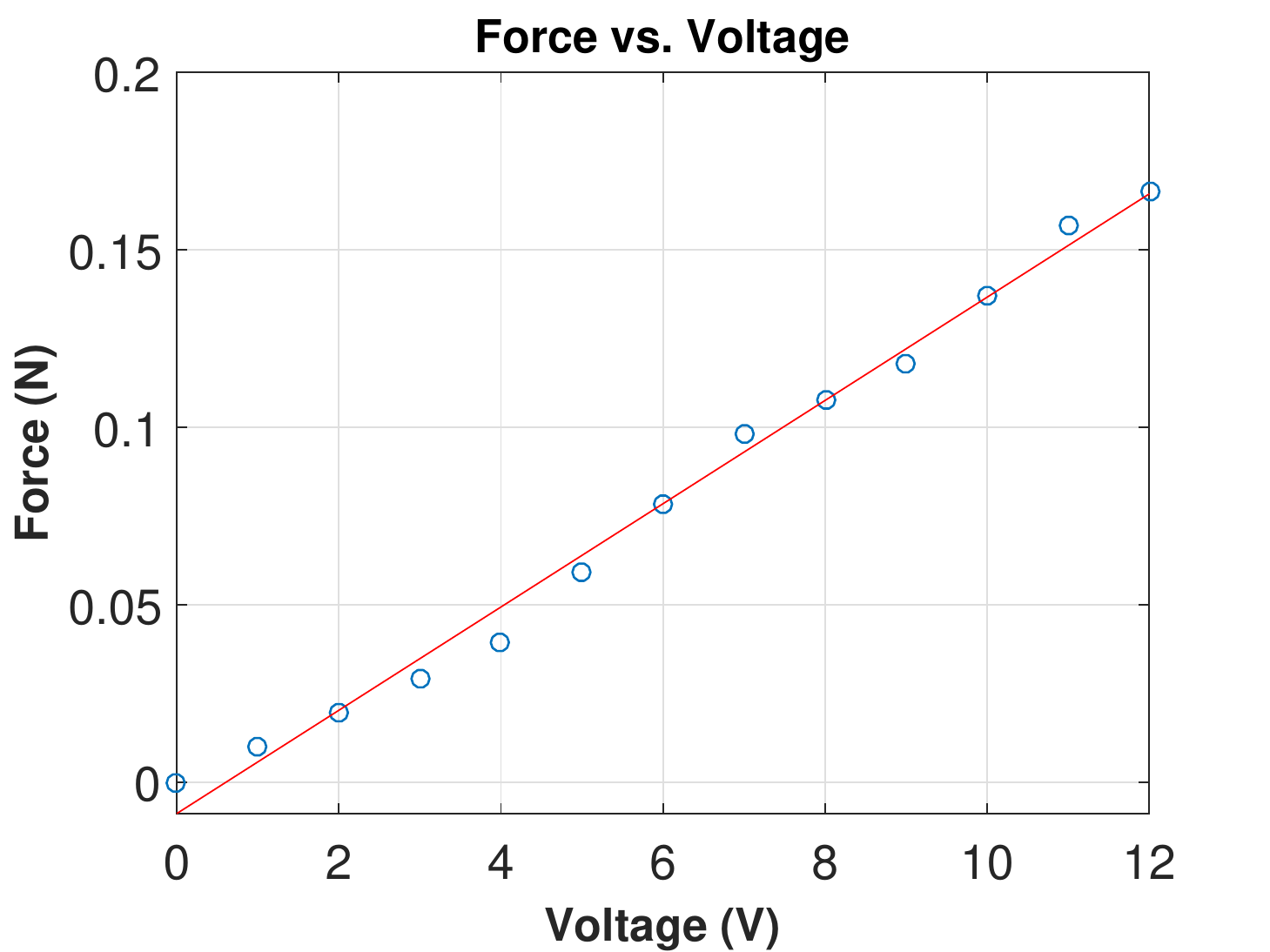}
\caption{Electromagnetic soft actuator generated force due to applying voltage to the coils caused interaction between the ferromagnetic core and magnetized coils.}
\label{fig:graph}
\vspace{-0.4cm}
\end{figure}

Two helical coils are embedded in silicone rubber and powered having a variable input between 0 and 12 Volts to control the force and position of the soft actuator. The helical coil provides the electromotive force of the soft actuator by the reaction of magnetic field to the current passing through it. The produced magnetic field causes the helical coil to react to magnetic field from a permanent magnet core fixed to the soft actuator’s frame, thereby provides expansion/contraction to the ESA. The spring linkage is the system that provides the elongation for the soft actuator as the electromotive force is axially applied.
Moreover, there are some electrical components and circuitry including: main power switch, power supply, voltage regulator, the Arduino Mega Microcontroller, H-bridge and the soft actuator to facilitate our design and test implementation.
 In order to obtain the generated force by ESA, force measurements were conducted at various voltages by means of a load cell. The soft actuator was fixed horizontally so that it could push against the load cell as voltage applied. The load cell is fixed using some clamps so that it stayed stationary. The force measurements were taken at various voltages and presented in Fig.~\ref{fig:graph}.
The relationship between the voltage and force is assumed to be linear for simplicity
\begin{equation}
F(t)\approx 0.0146 V(t) -0.0088.
\label{equation:equ1}
\end{equation}
This experimental part in the paper is needed for any ESA used to obtain the slop of the above curve. The verification of the relationship between the input force and voltage is essential in yielding a tractable computational framework instead of a nonlinear model that would further complicate the analysis and design of robust control methods with disturbances. The next section presents an aggregate state-space model for networked ESAs that extends to networked soft-body actuators.
\section{State-Space Modeling for ExoMuscles}~\label{sec:ModelingSS}
The scalability of the actuator allows us to exert greater force from a brace with specific dimensions by embedding  more smaller soft actuator cells in a constant volume of the brace.
Our proposed structure for the ExoMuscle is a matrix of several ESA actuators. Every single actuator is generating a partial force by means of applying  voltage.  As a result, the total generated force of the array also is controllable by the voltage. 
Our suggested model resembles the structure and function of human muscle components. We consider the actuator as building blocks in developing simulated artificial muscle fiber. Fig.~\ref{fig:musc} illustrate that a human skeletal muscle is composed of several longitudinal muscle fibers. Each muscle fiber can then be further broken down into smaller myofibers, which themselves are serial chains of sarcomeres. Each sarcomere is composed of parallel arrangement of muscle actuation units Actin and Myosin. 

\begin{figure} [t]
	\centering 
\includegraphics[scale=0.31]{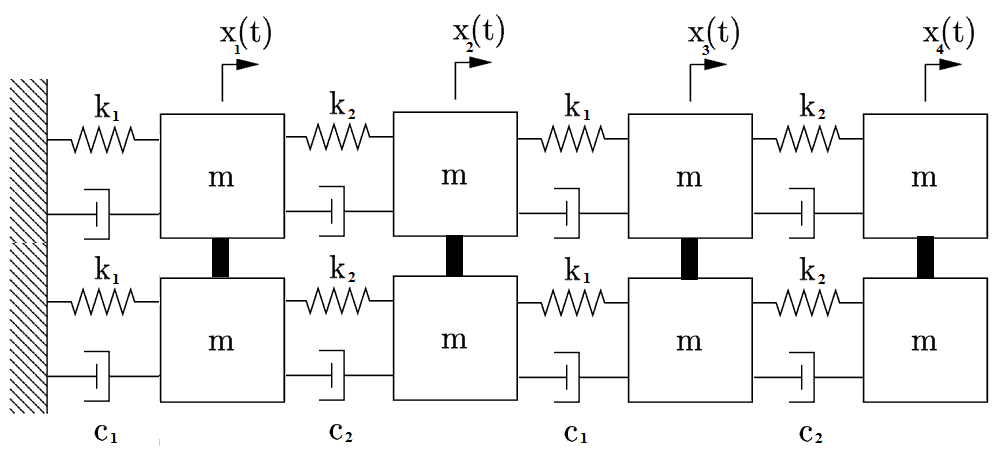}	
	\caption{Network with limited number of ESAs Forming ExoMuscles.}
	\label{fig:muscles}
\vspace{-0.45cm}	
\end{figure}

Following the same fashion, we first built the sarcomere part of our model by arranging bunch of soft actuators in parallel. Then, by connecting the modeled artificial sarcomeres in series and through silicone spring linkage which also modeled by spring and damper, we constructed an array representing artificial fibers. It is notable that the ExoMuscles would be the combination of artificial fibers in different arrangements and textures and would be used as active braces for rehabilitation and assistive application. To achieve the analytical model of the proposed exo-muscle, firstly, we model each ESA with the typical system of mass, spring damper as shown in Fig.~\ref{fig:muscles}.
Each soft actuator can be modeled by two masses ($m$) representing two conductive coils, one spring ($k_1$) and one damper ($c_1$) simulating the behavior of spring linkage. Each actuator connected to the next one  through another spring ($k_2$) and damper ($c_2$). Fig.~\ref{fig:muscles} depicts the schematic mass-spring-damper model of a $2$-by-$4$ matrix of mass-spring-damper networks. In each column of the matrix, the two masses of each actuator are attached to the subsequent ones. This structure resembles the actual muscle mechanism since there are columns of myosin and actin---akin to our proposed configuration. The motion dynamics of the ExoMuscle are
\begin{equation}
\dot{\m x}(t) = \m A \m x(t) + \m B_u \m u(t) + \m B_d \m d(t),
\end{equation}
where $\m x(t) \in \mbb{R}^{2rc=n_x}$ is the state-vector collecting all the deflection and velocities of each mass-spring subsystem; $\m u(t) \in \mbb{R}^{rc=n_u}$ collects all the input forces (or current injections) of all ESAs in the ExoMuscle; $\m d(t)\in \mbb{R}^{rc=n_d}$ includes the disturbances from external forces; $\m A, \m B_u$ and $\m B_d$ are a function of the parameters of the ExoMuscles, as well as the configuration which we discussed in the previous sections.  In short, the ExoMuscle has $n_x$ states, $n_u$ control inputs ($N=n_u/2$ ESAs), and $n_d$ external disturbances that are naturally matched by the control input.
We give a concrete example of a network of two series ESAs; the dynamics of ESAs can be written as
\vspace{-0.5cm}

{\small\begin{subequations}\label{equ:networkOfMasses}
	\begin{align}
	m\ddot{x}_1(t)&=-(k_1+k_2)x_1(t)-(c_1+c_2)\dot{x}_1(t)\notag \\
	&  +k_1x_2(t)+c_1\dot{x}_2(t)+F_1(t)+ F^{ex}_1(t) 
	\label{equation:equ2a} \\
	m\ddot{x}_2(t)&=k_1x_1(t)-(k_1+k_2)x_2(t)+k_2x_3(t)+ c_1\dot{x}_1(t)\notag \\&  \hspace{-0.011cm}-(c_1+c_2)\dot{x}_2(t)+c_2\dot{x}_3(t)+F_1(t)+ F^{ex}_1(t)
 \label{equation:equ2b} \\
	m\ddot{x}_3(t)&=k_2x_2(t)-(k_1+k_2)x_3(t)+k_1x_4(t)+ c_2\dot{x}_2(t)\notag \\&  -(c_1+c_2)\dot{x}_3(t)+c_2\dot{x}_4(t)+F_2(t)+ F^{ex}_2(t)
 \label{equation:equ2c} \\
	m\ddot{x}_4(t)&=k_1x_3(t)-(k_1+k_2)x_4(t)+k_2x_5(t)+ c_1\dot{x}_3(t)\notag \\&  -(c_1+c_2)\dot{x}_4(t)+c_2\dot{x}_5(t)+F_2(t)+ F^{ex}_2(t)\label{equation:equ2d} 
		\end{align}
\end{subequations}}
where $k_1$ and $c_1$ are the stiffness and damping coefficients between two masses of one actuator and $k_2$ and $c_2$ are the stiffness and damping coefficients between two ESAs; $x_{i}(t)$ denote the deflection; $F_{1,2}(t)$ are the generated forces via applying voltage to the first and second coils of first actuator (that we wish to solve for); $F^{ex}_{1,2}(t)$ are the unknown, yet matched external forces and disturbances.  

The next section investigates novel methods that optimally determine the combination of time-varying activated actuators in addition to the optimal input forces (input voltages) for all actuators, while considering worst-case disturbances and uncertainties due to the nature of the ESA networks.
	{It is noteworthy to mention that there are two sources of disturbance in our model. Firstly, the approximation of actual force-voltage graph by a straight line. Secondly, the constant portion of \eqref{equation:equ1} considered as a kind of disturbance inasmuch as the fact that practically in the absence of applied voltage there is no exerted force from soft actuator. }

\section{Robust Control and Actuator Selection }~\label{sec:RobustControl}
The previous sections focus on the modeling and the design of output-force maximizing methods for networked soft bodies and ExoMuscles, thereby achieving the high-level objectives of soft robotics. The objective of the remaining of the paper is two-fold. First, to investigate combinatorial optimal control methods that guide the selection of specific actuators for different time-periods under uncertainty and actuator logistic constraints. Second, to simultaneously design local control laws for individual actuators while taking into account bounds on the input energy (maximum input current/voltage) and worst-case disturbance scenarios.  The  objectives/challenges are coupled, and the theme of these methods is a set of tractable computational methods that explore solutions to the two challenges.  

\subsection{Robust Control Law and Actuator Selection}
First, we define $\m \Gamma \in \mathbb{R}^{n_u \times n_u}$ to be a diagonal matrix with binary variables $\Gamma_i$ that select the optimal actuators, where $\Gamma_i=1$ if the $i$th actuator is selected and $0$ otherwise. The ExoMuscles dynamics derived in Section~\ref{sec:ModelingSS} with the time-varying actuator selection in time-period $j$ can be written as
\vspace{-0.5cm}
\begin{subequations}\label{equ:SSDynamics}
\begin{align}
\m {{\dot x}}(t) &={\m A}\m { x}(t)+ {\m B}_u \m \Gamma^j  \m { u}(t) + \m B_d \m {  d}(t),\\
\m p(t) &=  \m C \m x(t) + \m D  \m u(t) .
\end{align}
\end{subequations}
The performance index $\m p(t)$ is a vector that quantifies the quantities to be minimized where $\m C$ and $\m D$ reflect the weights---akin to the linear quadratic regulator cost function. The need for designing a robust control law for various time periods is due to: (a) the need to deactivate actuators to avoiding overheating, (b) the possibility of fatigue which impacts the performance of actuators, and (c) potential damages to actuators which means that these have to be deactivated permanently. Other logistic constraints also justify the need for a time-varying actuator selection with robust control. We solve for variable $\m \Gamma^j$ via this framework which is reflected via the following optimization routine. 
\begin{subequations}~\label{equ:HLOpt}
	\begin{eqnarray}
	\minimize_{\m K^j, \m \Gamma^j } && \sum_{j=1}^{T} \frac{ \| \m p(t) \|_\infty   }{\| \m d(t) \|_{\infty} } + \alpha_{\Gamma} \mathrm{trace}(\m \Gamma^j) \\
	\subjectto && 	\eqref{equ:SSDynamics} , \m u(t) = \m K^j \m x(t),\; \|  \m u(t)\| \leq u_{\max}\\
	&& \;\; \m \Gamma_i^j \in \{0,1\}^N,\;\;\m \Gamma^j \in \mathcal{A}^j.
	\end{eqnarray}
\end{subequations}
Problem~\eqref{equ:HLOpt} minimizes the impact of the worst-case disturbance $\| \m d(t) \|_{\infty}$ on the performance index, in addition to the weighted number of activated actuators (weight is $\alpha_{\Gamma}$). The constraints are the uncertain physics-based dynamics, the state-feedback controller through time-varying gain $\m K^j$ which is fortunately possible through soft sensors and dynamic state estimation methods, input-voltage budget constraints $u_{\max}$, binary constraints on the actuator selection, and logistic constraints $\m \Gamma^j \in \mathcal{A}^j$ as described above. {We do not elaborate on soft sensors and the observability of the dynamic system here, but we assume that the states are available. Since soft sensors cannot measure all the states of the ExoMuscle, a simple state estimator can be used to obtain $\hat{\m x}(t)$, the estimate of $\m x(t)$.} The input constraints can be considered on the maximum available input voltage, whereas the logistic constraints represent simple actuator-based rules for various time-periods. The optimization problem~\eqref{equ:HLOpt} is a very challenging combinatorial, non-convex optimization problem. 

\begin{figure*}
	\begin{subequations}~\label{equ:SASLINFOPT}
		\begin{align}
	f^*=	\minimize_{\{\m P, Y, \mu_0,\mu_1,\mu_2,\alpha\}^j} & \;\;\; \sum_{j=1}^{T} \mu_0^j \mu_1^j + \mu_2^j + \alpha_{\Gamma}\mathrm{trace}\left(\m \Gamma^j\right) \\
	\subjectto \hspace{0.6cm}& 	\begin{bmatrix}
		\m P^j\m A^{\top}+\m A\m P^j+\alpha \m P^j \\ -\m Y^{j\top} \m \Gamma^{j\top}\m B_u^{\top}-\m B_u\m \Gamma^j\m Y^j& \m B_d \\
		\m B_d^{\top} & -\alpha \mu_0^j \m  I
		\end{bmatrix} \preceq 0 ,\;\; \begin{bmatrix}
		-\mu_1^j\m P^j & \m O & \m P^j\m C^{\top}+\m Y^{j\top}\m D^{\top}\\ \m O & -\mu_2^j\m I & \m O \\ \m C\m P^j +\m D \m Y^j& \m O & -\m I
		\end{bmatrix}
		\preceq 0 ~\label{equ:SASLINFOPT-1x}\\
		&\bmat{-\mu_0^j\rho^2 & \m x_0^{\top} \\ \m x_0 & -\m P^j} \preceq 0, \;\; \bmat{-\frac{u_{\max}^2}{\rho^2}\m P^j & \mu_0\m Y^{j\top}\\ \mu_0^j\m Y^j & -\mu_0^j \m I} \preceq 0 ,\;\m \Gamma_i^j \in \{0,1\}^N,\; \m \Gamma^j \in \mathcal{A}^j,\; j=1,\ldots,T.~\label{equ:SASLINFOPT-2x}
		\end{align}
	\end{subequations}
	\hrulefill
\end{figure*}
\subsection{Problem Formulation }
Using the notion of $\mathcal{L}_{\infty}$-stability of LTI systems under disturbances from~\cite{pancake2000d}, we can show that Problem~\eqref{equ:HLOpt} can be written as an optimization problem with mixed-integer nonlinear matrix inequalities (MI-NMI).  Assuming that the initial conditions are $\m x_0$ and that $\| \m d(t) \|_{\infty} \leq \rho$, then, if there exist real matrices $\m P=\m P^{\top}\succ 0$ and $\m Y$, binary diagonal matrix $\m \Gamma$, and  scalars $\{\mu_0, \mu_1, \mu_2 \} > 0$ for all time-periods that are the solution to optimization problem~\eqref{equ:SASLINFOPT} (given in the next page) with optimal value $f^*$,
then the time-varying, adaptive feedback controller $\m u(t)=\m K^j \m x(t)$ with $\m K^j = -\m Y^j(\m P^j)^{-1}$ guarantees that 
$$\Vert \m p(t) \Vert_2 \leq \mu \rho,\;\; \mu = \max\left(\sqrt{\mu_0^j\mu_1^j+\mu_2^j}\right),\;\; \forall j$$ and that the closed loop system with unknown inputs with the selected actuators is $\mathcal{L}_{\infty}$-stable with performance level\footnote{The reader is referred to~\cite{pancake2000d} for more the definition of $\mc{L}_{\infty}$ stability, but in short this definition entails that the states of the linear system under disturbances are guaranteed to be bounded in a tube centered at the origin of radius $\mu$.}  $\mu$ with minimal number of actuators. 
We have recently analytically proved that~\eqref{equ:SASLINFOPT} is in fact an accurate representation of~\eqref{equ:HLOpt}; the proof is not presented in here, and left for an extended version of this manuscript\footnote{The interested reader can contact the corresponding author for more information related to the derivation.}. This formulation implies that regardless of the disturbance that the system is subject to, then the performance index remains within a neighborhood of the origin (or the final desired state) with the the best performance index $\mu$. 

Unfortunately, the optimization routine \eqref{equ:SASLINFOPT} is challenging to solve due to the MI-NMIs. Section~\ref{sec:EfficientMethods} outlines efficient computational methods to bound the optimal value $f^*$ of~\eqref{equ:SASLINFOPT}.
\section{Methods to Bound the Solution of~\eqref{equ:SASLINFOPT}}~\label{sec:EfficientMethods}
This section is dedicated to develop numerical computing methods that, in general, solve nonconvex optimization problems with MI-NMIs and in specific, robust control with actuator selection for uncertain soft-body dynamics~\eqref{equ:SSDynamics}. 

\subsection{Approach 1: Relaxing The Integer Constraints} The first approach is based on relaxing the integer constraints $\m \Gamma \in \{0,1\}$ to $\m \Gamma \in [0,1]$. This approach is common in various problems where integer variables appear in nonlinear and linear optimization~\cite{duran1986outer,bussieck2003mixed,boyd2004convex}. Later in this section, we discuss a simple approach to recover the integer variables from the real, continuous solutions.  Given this relaxation, and after removing the dependence on time-period $j$ simplicity, Problem~\eqref{equ:SASLINFOPT} becomes nonconvex problem with bilinear matrix inequalities (BMI) that can be written as
\begin{subequations}~\label{equ:SASLINLBUx}
\begin{align}
L^* = \minimize & \;\;\; \mu_0  \mu_1  + \mu_2  + \alpha_{\Gamma} \mathrm{trace}(\m \Gamma ) \\
	\subjectto &	 \;\; ~\eqref{equ:SASLINFOPT-1x}, \eqref{equ:SASLINFOPT-2x}, \;\; \m \Gamma \in [0,1],\;\; \m \Gamma  \in \mathcal{A},
\end{align}
\end{subequations}
where $L^* \leq f^*$ due to the integer relaxation and hence the expansion of the feasible space.
BMIs appear in various optimal control and state estimation problems in dynamic systems~\cite{boyd1994linear,vanantwerp2000tutorial}. Various methods have investigated upper bounds to optimization problems with BMIs; see~\cite{dinh2012combining}.  However, these methods have a widely acknowledged limitation: they offer no intuition or direction on recovering the optimal solution (unless costly, inefficient global optimization is used) to nonconvex problems with BMIs, that is $L^*$ in~\eqref{equ:SASLINFOPT}. 


\subsection{Upper Bound on $L^*$} 

Here, we present an approach based on successive convex approximations (SCA) to obtain an upper bound $L_u^*$ on $L^*$. Before discussing the SCA, we note that~\eqref{equ:SASLINFOPT} has the following BMI terms: $-\m B_u \m \Gamma \m Y-\m Y^{\top} \m \Gamma^{\top}\m B_u^{\top}$, $\mu_1 \m P$, $\mu_0 \m Y$, in addition to this nonconvex bilinear part of the objective function $\mu_0\mu_1$. We show how the SCA works for the most important bilinearity $-\m B_u \m \Gamma \m Y-\m Y^{\top} \m \Gamma^{\top}\m B_u^{\top}$ involving the actuator variables, and leave the other BMIs for brevity. 

First, notice that the aforementioned bilinear term can be written as
\begin{eqnarray*}
	-\mB_u  \m\Gamma  \m Y  -\m Y^{\top} \m\Gamma  \mB_u^{\top} &= \dfrac{1}{2}  \underbrace{\left( \mB_u  \m\Gamma  - \m Y^{\top}\right) \left(\mB_u  \m\Gamma  - \m Y^{\top}\right)^{\top}}_{\mc{P}(\m \Gamma, \m Y)}- \\
	&\dfrac{1}{2}\underbrace{\left( \mB_u  \m\Gamma  +\m Y^{\top}\right) \left(\mB_u  \m\Gamma  + \m Y^{\top}\right)^{\top}}_{\mathcal{H} (\m\Gamma ,\m Y )} .
	\label{eq:bilinDC}
	\end{eqnarray*}
The term $\mc{P}(\m \Gamma, \m Y)$ is convex in $\m Y $ and $\m\Gamma $, while $\mathcal{H} (\m\Gamma ,\m Y )$
is concave in $\m Y $ and $\m\Gamma $, and hence its first-order Taylor approximation is a global over-estimator. Let $\m\Gamma _0, \m Y_0 $ be the linearization point, and let $\mathcal{H}_{\mathrm{l}} (\m\Gamma ,\m Y ;\m\Gamma _0, \m Y_0 )$ denote the linearization of $\mathcal{H} (\m\Gamma ,\m Y )$ at the point $(\m\Gamma _0, \m Y_0 )$. It holds that 
$\mathcal{H} (\m\Gamma ,\m Y ) \preceq \mathcal{H}_{\mathrm{l}} (\m\Gamma ,\m Y ;\m\Gamma _0, \m Y_0 )$
for all $\m\Gamma _0, \m Y_0 $  and $\m\Gamma , \m Y $. The initialization point can be chosen as $\m \Gamma_0 = \m I_{n_{u}}$ (that is, all ESAs are activated). 	The bilinear terms can now be upper bounded and approximated via
\begin{align}
 \bmat{\mA \m P +\m P \mA^{\top}+\alpha\m P  \\-\mB_u  \m\Gamma  \m Y  -\m Y^{\top} \m\Gamma  \mB_u^{\top} &\mB_d  \\\mB_d^{\top} & -\alpha\mu_0\mI} 
&\preceq\notag  \\
& \hspace{-4.95cm}\bmat{\mA \m P +\m P \mA^{\top}+\alpha\m P    \\+0.5 ( \mathcal{H}_{\mathrm{l}}  + \mc{P}) &\mB_d  \\\mB_d^{\top} & -\alpha\mu_0 \mI} = \mc{C}_1(\m P, \m \Gamma, \m Y,\mu_0).
\label{eq:Cineq}
\end{align}
Notice that the RHS of~\eqref{eq:Cineq} is indeed an LMI in terms of $\m P, \m \Gamma, \m Y,$ and $\mu_0$. Hence $\mc{C}_1(\cdot)$ is a convex constraint. This approach can be applied to the other bilinearities and the objective function, leading to the approximation of~\eqref{equ:SASLINLBUx} to a convex problem
\begin{subequations}~\label{equ:SASLINLBU}
\begin{align}
{L}_u^* = \min & \;\;\; \mc{C}_o(\mu_0,\mu_1,\mu_2) + \alpha_{\Gamma} \trace(\m \Gamma ) \\
\st  &\;\;  \mc{C}(\m P, \m \Gamma, \m Y,\mu_0,\mu_1), \;\; \m \Gamma_i  \in [0,1],\;\; \m \Gamma  \in \mathcal{A},
\end{align}
\end{subequations}
where $\mc{C}(\cdot)$ collects all the convex approximations of the bilinear terms of the constraints in~\eqref{equ:SASLINLBUx}, while $\mc{C}_o(\cdot)$ \textit{convexifies} the nonconvex terms in the objective function of~\eqref{equ:SASLINLBUx}. {The explicit form of \eqref{equ:SASLINLBUx} with all the linear matrix inequalities and constraints is not provided in this paper due to the lack of space. However, following the convex approximation of the first constraint, other approximations can similarly be derived. } This problem is solved iteratively, given the linearization points, until a certain stopping criteria is satisfied as described in Algorithm~\ref{algorithm:SCA}---similar to all SCAs. It is noteworthy to mention that such approximations yield desirable convergence properties as discussed in our recent work~\cite{taha2017time}. 
	\begin{algorithm}[H]
				\caption{Solving SCA for \eqref{equ:SASLINLBU}.}
			\label{algorithm:SCA}
			\begin{algorithmic}
				
				\STATE \textbf{initialize} the bilinear terms, $k=0$
				\STATE \textbf{input:} $\mathrm{MaxIter},\mathrm{tol}$ 
				\WHILE {$k < \mathrm{MaxIter} $}
				\STATE Solve~\eqref{equ:SASLINLBU} 
				\IF{ $|L_{u_{k}}^*-L_{u_{k-1}}^*| < \mathrm{tol}$ }
				\STATE \textbf{break}
				\ELSE 
				\STATE $k\leftarrow k+1$
				\ENDIF		\ENDWHILE
			\end{algorithmic}
		\end{algorithm}	
\vspace{-0.7cm}

\subsection{Recovering the Integer Variables}\label{sec:slicing}
The solutions above generate real, continuous values for the actuator selection decision variable $\Gamma_i \in [0,1]$. To obtain the binary selection, the real actuator selection from the optimal solution of~\eqref{equ:SASLINLBU} and Algorithm~\ref{algorithm:SCA} can be ranked in decreasing order. Then, actuators are added up until a closed loop system metric  is satisfied. Alternatively, all actuators with $\Gamma_i^{\mathrm{real}} \geq 0.5$ can be activated. This approach yields effective results in highly nonconvex mixed-integer nonlinear programs~\cite{belotti2013mixed,lee2011mixed}. 
Any solution that we obtain from this approach yields an upper bound $U^*$ on $f^*$ since the selection will be feasible for the original problem~\eqref{equ:SASLINFOPT}. Therefore, Approach 1 yields tight lower and upper bounds on the optimal solution of~\eqref{equ:SASLINFOPT}: $L^* \leq f^* \leq U^*$. 
\section{Approach 2: Keeping the Integer Variables}~\label{sec:BM} As a departure from relaxing the integer constraints, we investigate methods that transform Problem~\eqref{equ:SASLINFOPT} to a mixed-integer semidefinite program (MI-SDP). This is useful as the recent studies~\cite{gally2017framework,tawarmalani2004global,lee2011mixed} have investigated and developed efficient, open-source solvers to solve MI-SDPs.
To see how that can be implemented, we examine the major mixed-integer nonlinear term $\m \Xi = \m \Gamma \m Y$. Due to the binary and diagonal nature of $\m \Gamma$, we can write $\Xi_{i,(a,b)}=
\Gamma_i Z_{i,(a,b)}$ if $\Gamma_i=1$ and $\Xi_{i,(a,b)}=
0$ if $\Gamma_i=0$ for $ i=1,\ldots,N,\, a=1,2$ (each actuator has two control inputs) and  $b=1,\ldots,n_x$, This relationship can be equivalently written as 
\begin{equation}
|\Xi_{i,(a,b)}-Z_{i,(a,b)}| \leq M (1-\Gamma_i), |\Xi_{i,(a,b)}| \leq M\Gamma_i
\label{eq:bilin-equiv-bigM}
\end{equation}
$M$ is a sufficiently large positive constant; matrix $\m \Xi$ replaces $\m \Gamma \m Y$  in the first matrix inequality~\eqref{equ:SASLINFOPT-1x} as follows 
\begin{equation}~\label{equ:txe}
\hspace{-0.34cm}{ \bmat{\mA\m P+\m P\mA^{\top}+\alpha\m P -\mB_u  \m\Xi -\m\Xi^{\top} \mB_u^{\top}&\mB_d \\\mB_d^{\top} & -\alpha\mu_0\mI}}\preceq 0. \end{equation}
The overall optimization problem using the Big-M method can be written as
	\begin{subequations}~\label{equ:bigMLINF}
	\begin{align}
	\min & \; \sum_{j=1}^{T} \mu_0^j \mu_1^j + \mu_2^j + \alpha_{\Gamma}\trace(\m \Gamma^j) \\
	\st & \begin{bmatrix}
	-\mu_1^j\m P^j & \m O & \m P^j\m C^{\top}+\m Y^{j\top}\m D^{\top}\\ \m O & -\mu_2^j\m I & \m O \\ \m C\m P^j +\m D \m Y^j& \m O & -\m I
	\end{bmatrix}
	\preceq 0 \\
	& \;\;\; \eqref{equ:SASLINFOPT-2x}, \eqref{eq:bilin-equiv-bigM},\eqref{equ:txe}.
	\end{align}
\end{subequations}
 Considering that $\alpha, \mu_0, \mu_1$ are all predetermined constants (not variables), then~\eqref{equ:bigMLINF} is in fact a MI-SDP with $\m\Gamma, \mu_2, \m Y, \m \Xi,$ and $\m P$ as the optimization variables. MI-SDPs can be solved via either off-the-shelf and industry-grade solvers or recently developed solver in~\cite{gally2017framework}. Many of these implementations use branch-and-bound to solve MI-SDPs. The user can also choose $\alpha, \mu_0, \mu_1$ to be variables, but that requires then performing the SCA with the MI-SDP to obtain solutions.


\section{Numerical Tests}\label{sec:numtests}
In this numerical tests, we demonstrate the applicability of the proposed method to the networked of artificial muscles consisting of 8 actuators, where each actuator has 6 rows of mass-spring systems. With this particular configuration, the state vector can be express as $\m x(t) = \bmat{\m x_p^{\top}(t) & \m x_v^{\top}(t)}^{\top}$  where the vectors $\m x_p(t)\in\mathbb{R}^{16}$ and $\m x_v(t)\in\mathbb{R}^{16}$ respectively collect the states of position and velocity of the whole system. The state space of the system is of the form
$
\dot{\m x}(t) = \m A \m x(t) + \m B_u \m u(t) + \m B_d \m d(t), \m p(t) =  \m C \m x(t) + \m D  \m u(t)$
where $\m A$ depicts the network of ESAs described in \eqref{equ:networkOfMasses} with parameters $m = 2.94\times 10^{-3}\,\mathrm{ kg}$, $k = 0.343\,\mathrm{N/m}$, $c = 1.75\times 10^{-16}\,\mathrm{N\cdot s/m}$ . The mass of each coil is obtained by measurement. Stiffness of the silicone rubber linkage is achievable according to its $100\%$  Modulus, average cross section and length using Hook's Law. Damping coefficient is assumed to be very small \cite{ecoflex}.
%

The remaining state-space matrices are specified as follows
\begin{align*}
	\m C = \bmat{0.1\times\m I_{16} & \m O_{16\times 16}},\quad \m D =  \bmat{0.01\times\m I_{8} \\ \m O_{8 \times 8}}.
\end{align*}
A sinusoidal signal is chosen to simulate the disturbance, which is represented as $\m d(t) = \mathrm{cos}(0.1t)$ with $\Vert\m d(t)\Vert_{\infty} = 2.8284$. The numerical tests are performed by assuming zero initial conditions within a timespan of $t_{\mathrm{span}} = [0,60] \sec$. For simplicity, we solve the problem in one time period only, that is $j = 1$ and leave the multi-time period numerical tests for extensions of this work. All numerical tests are performed using \iffalse MATLAB R2016b running on 64-bit Windows 10 with 3.4GHz Intel Core i7-6700 CPU and 16 GB of RAM \else MATLAB R2017b running on a 64-bit Windows 10 with 2.5GHz Intel Core i7-6500U CPU and 8 GB of RAM\fi. The details of the numerical tests are grouped into three following sections: (Scenario A) We show the results for the $\mc{L}_{\infty}$ control problem without actuator selection, to test the robustness of the system with a full set of 8 actuators; (Scenario B) We solve the SCA followed by the slicing algorithm discussed in Section~\ref{sec:EfficientMethods}; (Scenario C) We solve the formulation based on MI-SDPs shown in Section~\ref{sec:BM}.  Scenarios B and C return a specific feedback gain matrix $\m K^*=\mY^*(\mP^{-1})^*$ and $\m\Gamma^*$ as the computed actuator selection. YALMIP~\cite{yalmip} and MOSEK~\cite{mosek} are used to solve all of the optimization problems. 
\subsection{Results for Different Optimization Methods}
We do not list all of the specific details and parameters in the implementation of the SCAs. However, all the codes used to run the numerical tests are available upon request by contacting the corresponding author. 
\subsubsection{Scenario A}
Here, we put our interest to find the best performance index of the system with full set of 8 actuators. Since the problem involves BMIs, we solve the SCA formulation of \eqref{equ:SASLINFOPT} by setting $\m \Gamma = \m I$. The initialization of SCA is performed by setting $u_{\max} = 250$ (which is much larger that what controllers require), $\alpha = 10^{-3}$, $\mu_0 = 30$, and $\mu_1 = 4$. Since the SCA requires a starting point that lies inside the relative interior of the feasible set, then we define a constant $\epsilon_1 = 10^{-4}$ such that all the LMIs in \eqref{equ:SASLINFOPT} are $\preceq -\epsilon_1 \m I$, which is common in these studies~\cite{dinh2012combining,taha2017time}. For the SCA iterations, we follow Algorithm~\ref{algorithm:SCA} with stopping criteria specified by maximum number of iterations, that is set to be $50$, and convergence of the objective function  defined by $| L_{u_{k}}^*-L_{u_{k-1}}^*| \leq 10^{-2}$.

Table \ref{tab:result1} illustrates the values achieved in this scenario for $\alpha$, $\mu=\sqrt{\mu_0\mu_1+\mu_2}$ which is the performance level of the closed loop system, the computational time $\Delta t(\mathrm{sec})$, and $\sum_i \Gamma_i$ (the total number of activated actuators). Fig. \ref{fig:norm_z}a depicts the system performance in terms of the norm of the disturbance signal $\m d(t)$, the state $\m x(t)$, and the performance vector $\m p(t)$. 
\subsubsection{Scenario B}
In this scenario, we use the same parameters as those in Scenario A. First, an optimal solution is computed from~\eqref{equ:SASLINLBU} and Algorithm~\ref{algorithm:SCA}. Next, based on the output of Algorithm~\ref{algorithm:SCA}, an integer solution is computed via a simple {slicing algorithm} as discussed in Section~\ref{sec:slicing}. The numerical results are shown in Scenario B of Table \ref{tab:result1}, whereas the comparison between performance index and disturbance is shown in Fig. \ref{fig:norm_z}-b.
\subsubsection{Scenario C}
\begin{figure*}
	\vspace{-0.4cm}
	\subfigure[]{\includegraphics[width=0.35\linewidth]{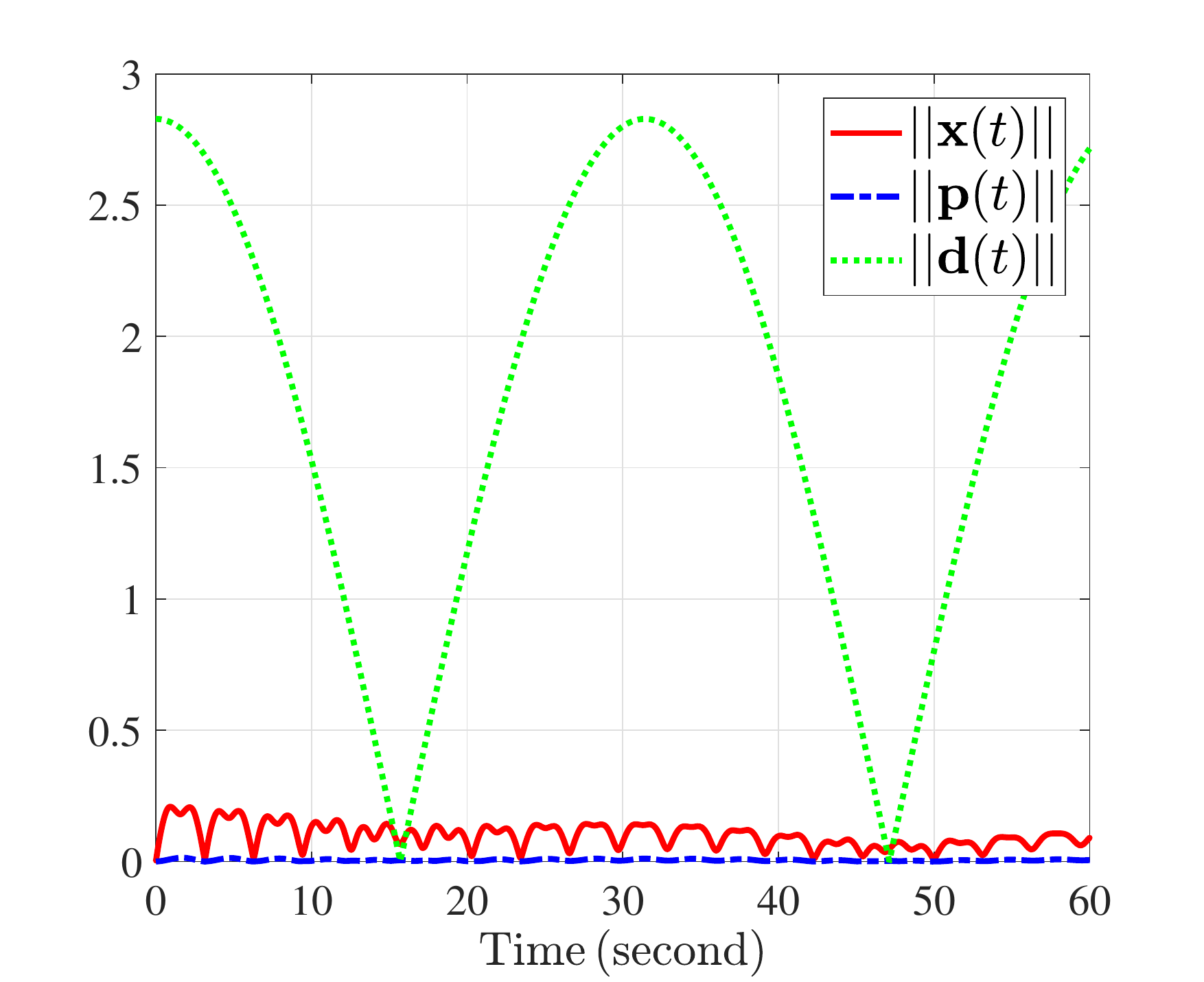}}\hspace{-0.3cm}
	\subfigure[]{\includegraphics[width=0.35\linewidth]{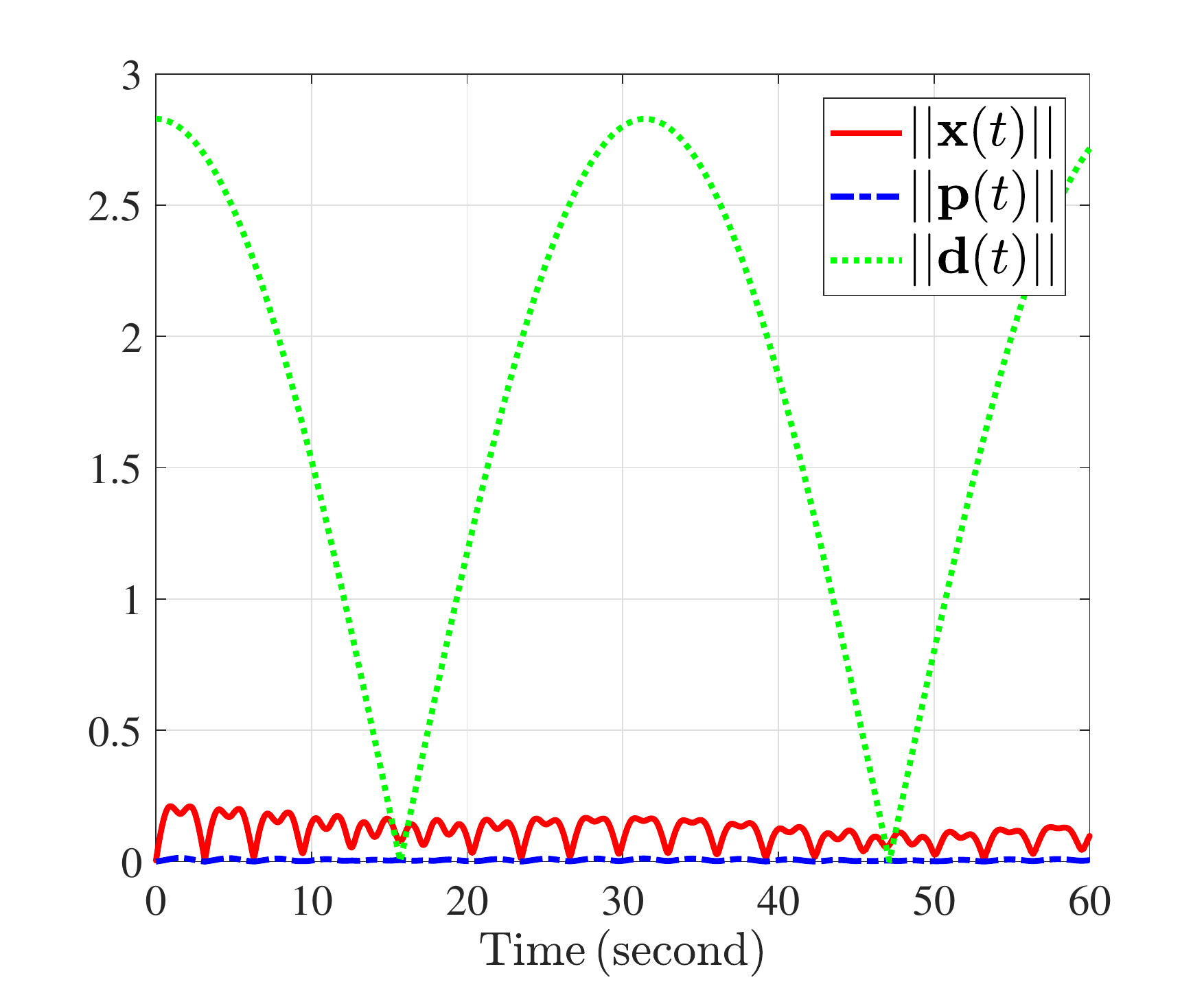}}\hspace{-0.3cm} \subfigure[]{\includegraphics[width=0.35\linewidth]{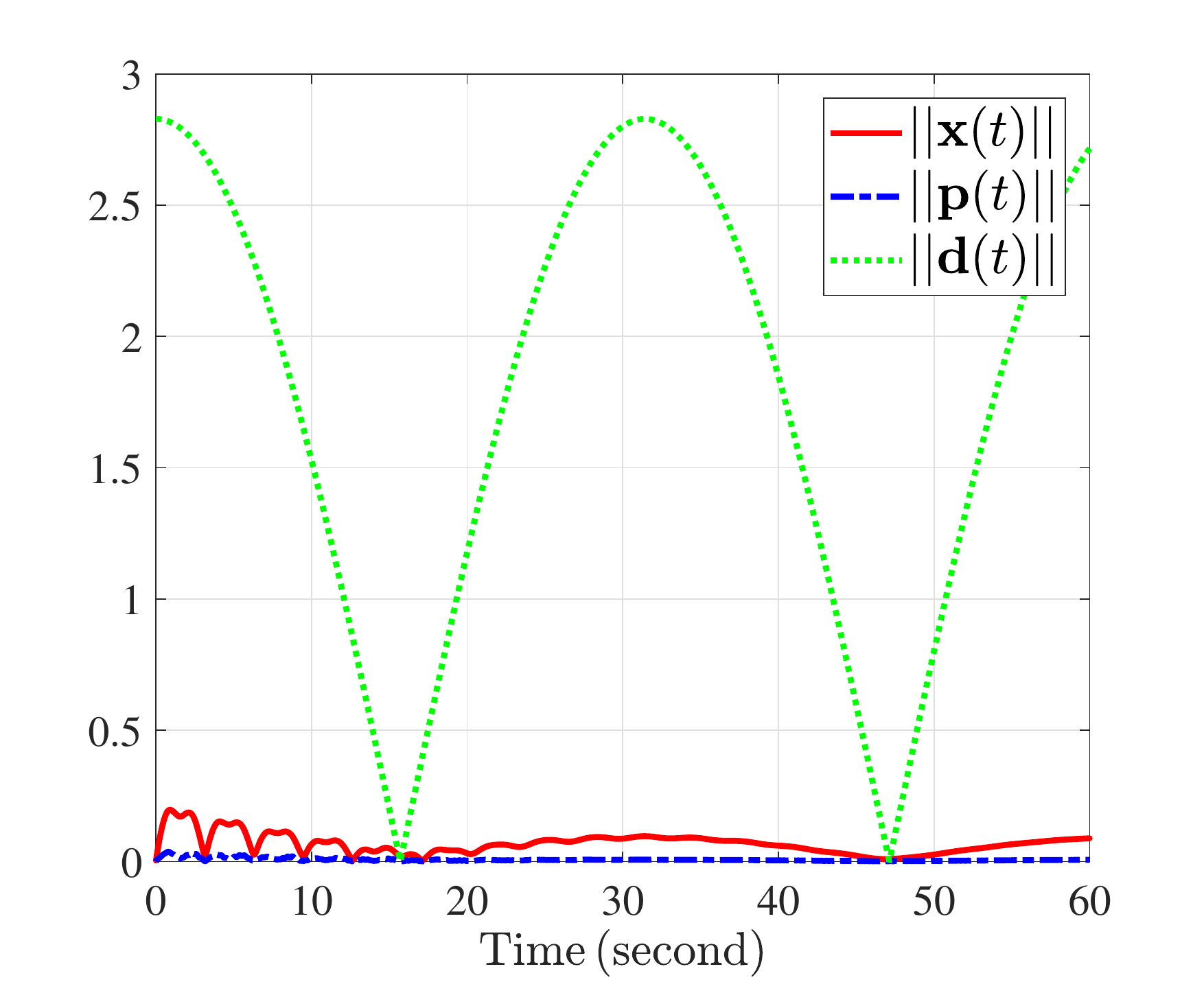}}\hspace{-0.3cm}
		\vspace{-0.4cm}
	\caption{Norm of $\m x(t)$, $\m p(t)$, and $\m d(t)$ for the (a) fully-actuated system (Scenario A), (b) the SCA Scenario B, and (c) the MI-SDP Scenario C. }
	\label{fig:norm_z}
	\vspace{-0.3cm}
\end{figure*}

In this particular test, the user can define $\alpha, \mu_0, \mu_1$ to be variables, but that requires then performing the SCA with the MI-SDP to obtain solutions. Alternatively,  we can predefine these variables to be constants. To that end, we use values similar to the solutions of Scenario B. The values are $\alpha = 0.1$ , $\mu_0 = 1$, and $\mu_1 = 0.4$, with big-M constant $M=10^3$; see \eqref{eq:bilin-equiv-bigM}. The MI-SDP is solved through YALMIP's branch-and-bound algorithm. The results of this test are shown in Table \ref{tab:result1} and Fig. \ref{fig:norm_z}-c. 

\begin{table}[t]
	\centering
	\caption{Numerical results for the three scenarios: (A) Fully actuated system, (B) Minimally actuated system via convex programming, (C) Minimally actuated system via the MI-SDP.}
	\begin{tabular}{|c|c|c|c|c|c|}
		\hline
		Scenario & {$\alpha$} & {$\mu$} & {$\sum_i\Gamma_i$} & {$\Delta t(s)$} & $\mathrm{Diag}(\m\Gamma)$ \\
		\hline \hline
		(A)& 0.169 & 0.44 &8     &58.3&$\lbrace 1,1,1,1,1,1,1,1\rbrace$ \\
		(B) & 0.104 & 0.62 &  3     & 189.9 &$\lbrace  0  ,   0  ,   0   ,  0   ,  1  ,   0  ,   1  ,   1\rbrace$\\
		(C) & 0.100 &0.63 & 2     & 74.4 & $\lbrace0  ,   0  ,   0  ,   1   ,  0 ,    0  ,   0   ,  1  \rbrace$ \\
		\hline
	\end{tabular}%
	\label{tab:result1}%
	\vspace{-0.4cm}
\end{table}%

\begin{figure} [h]
	\vspace{-0.2cm}
	\centering	\includegraphics[scale=0.4]{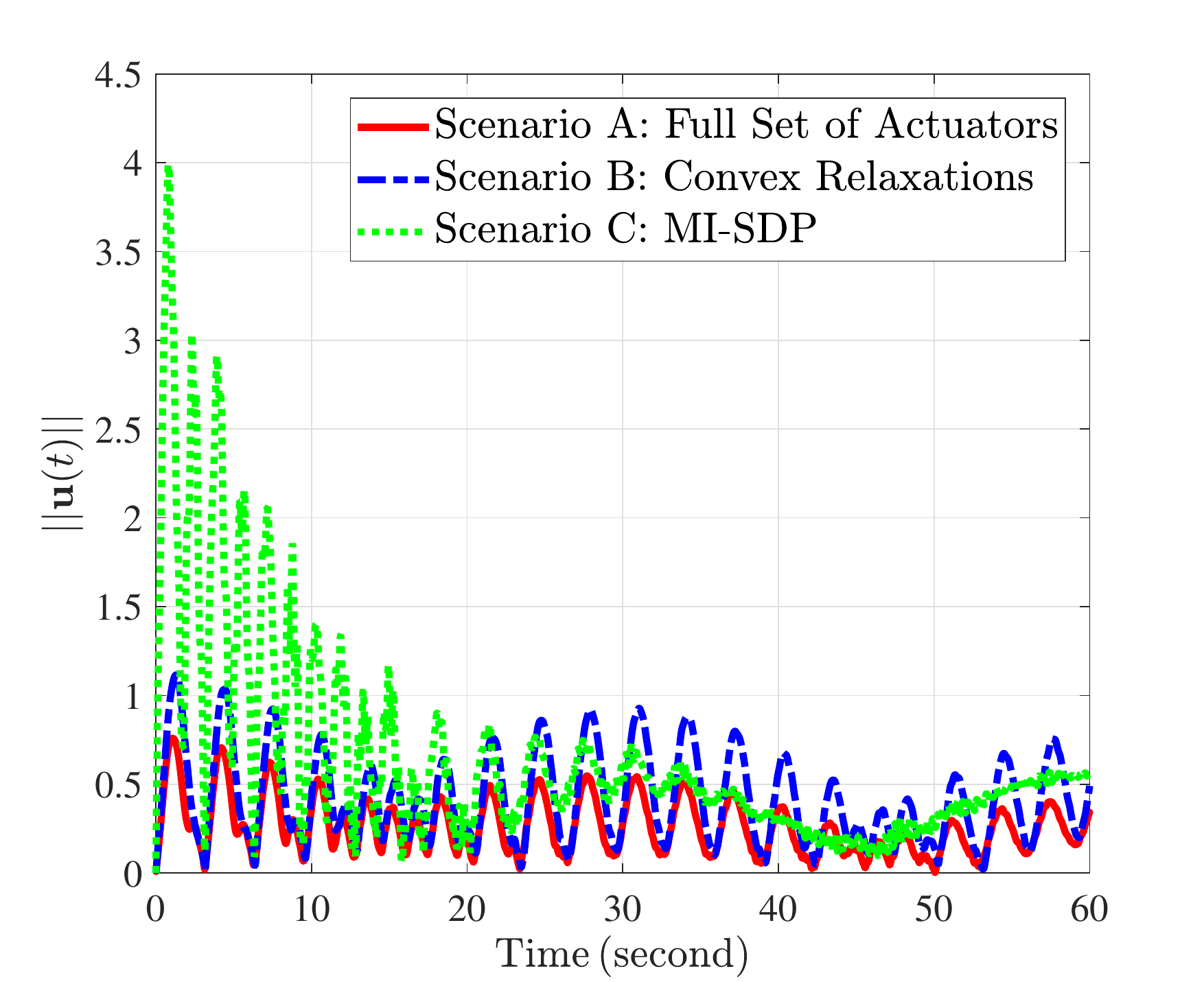}
		\caption{Comparison of the norm of $\m u(t)$ for Scenarios A, B, and C.}
		\label{fig:norm_u}
	\vspace{-0.24cm}
\end{figure}

\begin{figure} [h]
	\vspace{-0.12cm}	
	\centering	\includegraphics[scale=0.4]{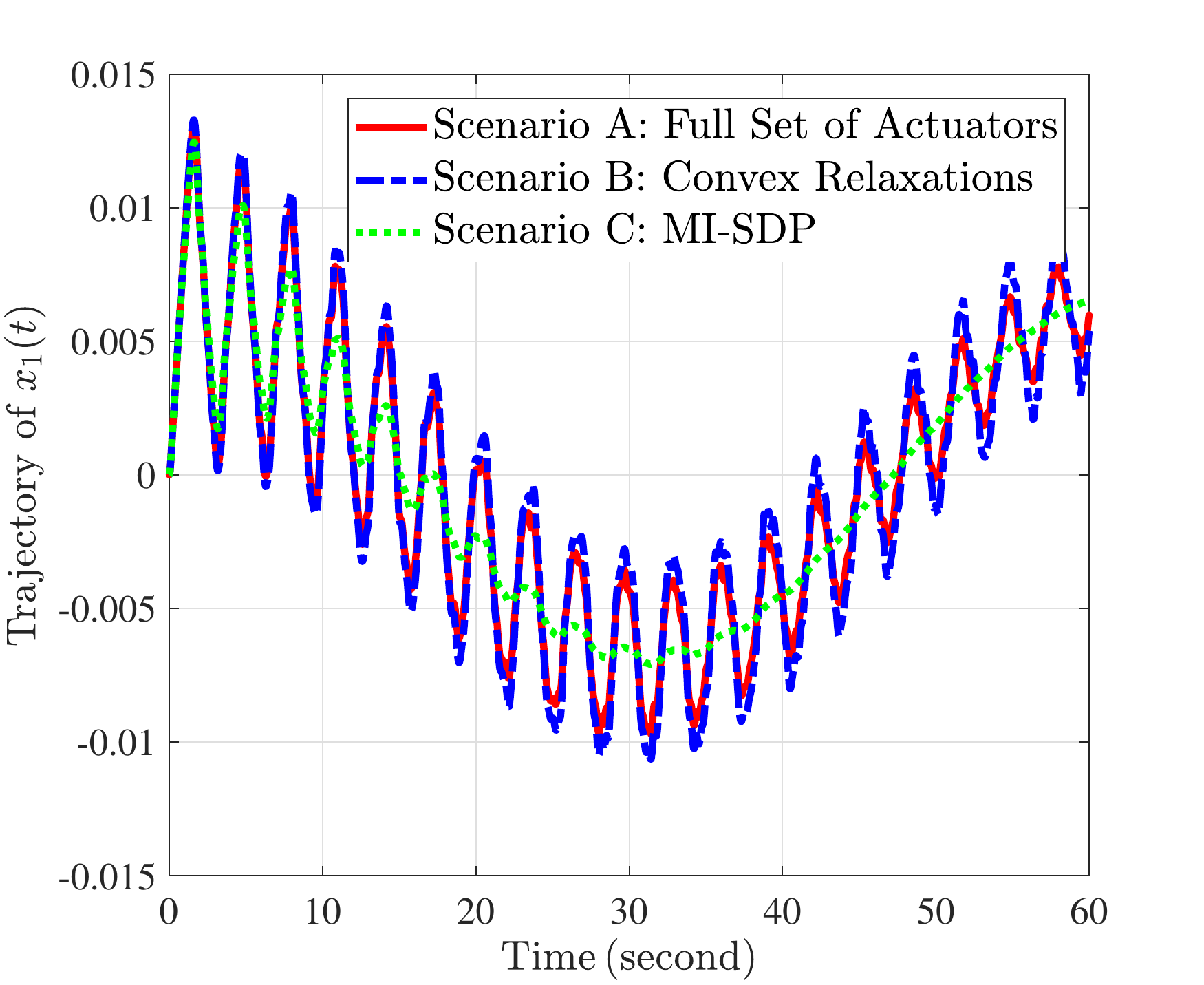}
		\caption{Comparison of the trajectory of $ x_1(t)$ for Scenarios A, B, and C.}
		\label{fig:state_trajectory}
	\vspace{-0.25cm}
\end{figure}

\subsection{Discussion and Insights}
The following {preliminary conclusions} are drawn from the experiments.
\begin{itemize}
	\item Scenario A uses all actuators. Consequently, it should provide the user with the best performance index $\mu$ in comparison with the other scenarios. This is corroborated by the numerical experiments; see Table~\ref{tab:result1}. 
	\item In Scenario B, a suboptimal solution of the actuator selection is given. In this scenario, the system only needs 3 actuators to be activated with performance index $\mu$ comparable to the one in Scenario A. The selection of actuator can be seen in Scenario B of Table~\ref{tab:result1}. This illustrates an interesting observation: \textit{similar performance index $\mu$ can be obtained with fewer number of actuators}.
	\item In Scenario C, the MI-SDP approach gives the least number of active actuators with computational time similar to that of Scenario A and performance index slightly worse than that of Scenario B; see $\Delta t(s)$ and $\mu$ in Table~\ref{tab:result1}. The specific actuator selection of actuator is shown in Table \ref{tab:result1}.
	\item The comparison of the input norm $||\m u(t)||$ which measures the energy of the control input (the voltage) is depicted in Fig. \ref{fig:norm_u} for the three scenarios. The results illustrate that the one with fewest number of actuators (Scenario C) requires much bigger input energy during the first 20 seconds in comparison with Scenario A or B that utilizes more active actuators. The control input norm for Scenarios A and B are comparable, although the latter tends to be bigger due to the utilization of fewer actuators. This presents another interesting observation: \textit{fewer actuators can be activated while not requiring much more input energy and also maintaining a comparable performance index. }
	\item Fig. \ref{fig:state_trajectory} illustrates the trajectory of the first state $x_1(t)$ for all scenarios. We can see that the damping of the oscillation of $x_1(t)$ for Scenario C is better than those of Scenario A or B. This happens because Scenario C uses more input energy to control the states, compared with the other scenarios.
\end{itemize}

\vspace{-0.5cm}
\section{Summary and Conclusions}\label{sec:conclusions}
The objective of this paper is to establish a framework that guides the design, analysis, and robust control of soft-body and ESA networks under uncertainty. A novel ESA is described, and a configuration of soft actuator matrix to resemble artificial muscle fiber is presented. A mathematical model which depicts the physical network is derived, considering the disturbances due to external forces and linearization errors as an integral part of this model. Combinatorial optimal control problems for uncertain dynamic ExoMuscles with actuator selection are formulated. To address the computational complexity, efficient computational routines are formulated based on convex programming and mixed-integer convex programming. Numerical tests show a promising simulated performance. Future work will focus on implementing the controller on an actual ExoMuscle testbed.
One of the limitations of the proposed networked ESAs is their resolution, which is highly dependent to the number of actuation units embedded in the system.  Compared to the real muscle composed of tremendous number of muscle actuation units, ExoMuscles are considered discrete and the resolution depends on the number and consequently the size of actuators. To achieve higher resolution we need to take advantage of high precision technologies such as accurate 3D printers. 

\vspace{-0.5cm}
\bibliographystyle{IEEEtran}	\bibliography{bibliography}

\end{document}

%% file: preambleT.tex
\usepackage{mathtools}
\usepackage{lipsum}
\usepackage{graphicx}
\usepackage{blindtext, graphicx}   
\usepackage{amsmath}   
\usepackage[tc]{titlepic}   
\usepackage{amssymb}   
\usepackage{amsfonts}   
\usepackage{lipsum}   
\usepackage{float}   
\usepackage{cuted}     
\usepackage{amsthm}   
\newtheoremstyle{case}{}{}{}{}{}{:}{ }{}
\usepackage{enumitem}
\setlist[itemize]{leftmargin=*}
\setlist[description]{leftmargin=*}

\usepackage{epsfig,color,amsmath,cite}
\usepackage{amsthm}

\usepackage{wrapfig}
\usepackage{bm}
\usepackage{epstopdf}
\usepackage{amssymb}
\usepackage{url}
\usepackage{enumitem}
\usepackage{multirow}
\usepackage{booktabs}
\usepackage{algorithm,algorithmic}	
\usepackage{comment}
\usepackage{cite}

\DeclareMathOperator{\trace}{trace}

\DeclareMathOperator*{\minimize}{minimize}

\DeclareMathOperator{\subjectto}{subject\ to}

\makeatother
\DeclareMathAlphabet\mathbfcal{OMS}{cmsy}{b}{n}

\newcommand{\mat}[1]{\boldsymbol{#1}}

\newcommand{\bmat}[1]{\begin{bmatrix} #1 \end{bmatrix}}

\providecommand{\mA}{\ensuremath{\mat{A}}}
\providecommand{\mB}{\ensuremath{\mat{B}}}

\providecommand{\mI}{\ensuremath{\mat{I}}}

\providecommand{\mP}{\ensuremath{\mat{P}}}

\providecommand{\mY}{\ensuremath{\mat{Y}}}

\newcommand{\st}{{\rm s.t.}}

\newcommand{\m}{\boldsymbol}
\allowdisplaybreaks[4]
\usepackage[colorlinks = true,
linkcolor = black,
urlcolor  = blue,
citecolor = blue,
anchorcolor = blue]{hyperref}

\newcommand{\mc}[1]{\mathcal{#1}}
\newcommand{\mbb}[1]{\mathbb{#1}}